\documentclass[12pt]{iopart}
\usepackage[colorlinks=true,bookmarks=false,citecolor=blue,urlcolor=blue]{hyperref} %pdflatex
\usepackage{amssymb}
\usepackage{amsfonts}
\usepackage{amsthm}
\usepackage{amsmath}
\usepackage{inputenc}
\usepackage{graphicx}
\usepackage{color}
\usepackage{braket}
\usepackage{mathptmx,courier,textcomp}

\usepackage{epstopdf}
\begin{document}

\title{Robust and scalable rf spectroscopy in first-order magnetic sensitive states at second-long coherence time}

\author{C.-H. Yeh$^{1,*}$, K. C. Grensemann$^{1}$, L. S. Dreissen$^{1}$, H. A. F\"urst$^{1,2}$, T. E. Mehlst\"aubler$^{1,2,3}$}

\address{$^{1}$ Physikalisch-Technische Bundesanstalt, Bundesallee 100, 38116 Braunschweig, Germany\\
$^{2}$ Institut f\"ur Quantenoptik, Leibniz Universit\"at Hannover, Welfengarten 1, 30167 Hannover, Germany\\
$^{3}$ Laboratorium f\"ur Nano- und Quantenengineering, Leibniz Universit\"at Hannover, Schneiderberg 39, 30167 Hannover, Germany}
\ead{$^{*}$chih-han.yeh@ptb.de}
\vspace{10pt}
%\begin{indented}
%\item[]December 2022
%\end{indented}

\begin{abstract}
Trapped-ion quantum sensors have become highly sensitive tools for the search of physics beyond the Standard Model. Recently, stringent tests of local Lorentz-invariance (LLI) have been conducted with precision spectroscopy in trapped ions~\cite{Pruttivarasin2015,Megidish2019,Sanner2019,Dreissen2022}. We here elaborate on robust radio-frequency composite-pulse spectroscopy at second long coherence times in the magnetic sublevels of the long-lived $^{2}F_{7/2}$ state of a trapped $^{172}$Yb$^{+}$ ion which is scalable to spatially extended multi-ion systems. We compare two Ramsey-type composite rf pulse sequences, a generalized spin-echo (GSE) sequence~\cite{Shaniv2018} and a sequence based on universal rotations with 10 rephasing pulses (UR10)~\cite{Genov2017} that decouple the energy levels from magnetic field noise, enabling robust and accurate spectroscopy. Both sequences are characterized theoretically and experimentally in the spin-$1/2$\ $^{2}S_{1/2}$ electronic ground state of $^{172}$Yb$^+$ and results show that the UR10 sequence is 38 (13) times more robust against pulse duration (frequency detuning) errors than the GSE sequence. We extend our simulations to the eight-level manifold of the $^2F_{7/2}$ state, which is highly sensitive to a possible violation of LLI, and show that the UR10 sequence can be used for high-fidelity Ramsey spectroscopy in noisy environments. The UR10 sequence is implemented experimentally in the $^2F_{7/2}$ manifold and a coherent signal of up to 2.5\,s is reached. In reference~\cite{Dreissen2022} we have implemented this sequence and used it to perform the most stringent test of LLI in the electron-photon sector to date with a single Yb$^{+}$ ion. Due to the high robustness of the UR10 sequence, it can be applied on larger ion crystals to improve tests of Lorentz symmetry further. We demonstrate that the sequence can also be used to extract the quadrupole moment of the meta-stable $^{2}F_{7/2}$ state, obtaining a value of $\Theta\,=\,-0.0298(38)\,ea^{2}_{0}$ which is in agreement with the value deduced from clock measurements~\cite{Lange2020coherent}.
\end{abstract}

%\maketitle

\section{Introduction}
Over the past decades, trapped-ion quantum sensors have become sensitive probes for physics beyond the Standard Model~\cite{Safronova2018}. Competitive tests of fundamental physical principles have been conducted with precise measurements of quantum mechanical resonances, see, e.g., references~\cite{Sanner2019,Acme2018,Kennedy2020}. Often, magnetic sensitive states, which strongly interact with the environment, need to be addressed spectroscopically. Decoherence from this unwanted interaction often hampers the interrogation time and, as a result, the resolution of such spectroscopic measurements. Methods to protect the system from environmental noise, developed in the field of quantum information science, are now being exploited for robust quantum sensing and more stringent tests of fundamental physics~\cite{Pruttivarasin2015,Hosten2016,Manovitz2019,Pedrozo2020}.

In trapped ions, decoherence-free entangled states have been used to suppress the influence of magnetic field noise~\cite{Roos2006designer,Kielpinski2001decoherence}, e.g., to test local Lorentz-invariance (LLI) with up to four entangled Ca$^{+}$ ions~\cite{Megidish2019} and to measure isotope shifts at the level of 10\,mHz in a pair of Sr$^{+}$ ions~\cite{Manovitz2019}. In states with finite lifetime, entangled states decohere faster than non-entangled ones, which may limit the interrogation time~\cite{Megidish2019}. On the other side, for states with extremely long lifetime as in ytterbium ~\cite{Lange2021}, the slow excitation rate leads to a decoherence of the entangled states already in the preparation stage. Another approach to protect the quantum system from magnetic field noise was invented by the nuclear magnetic resonance (NMR) community where clever spin-echo like sequences have been invented to decouple the quantum system from noise~\cite{Hahn1950spin,Slichter2013principles}. In the atomic physics community, this method has been adopted and pulsed electromagnetic fields are used to engineer a robust system~\cite{Viola1998,Viola1999,Wang2021}. A proof-of-principle implementation of a generalized spin-echo (GSE) Ramsey sequence in the $^2D_{5/2}$ state in Sr$^{+}$, to reach coherence times of up to 30\,ms with 110 rephasing pulses~\cite{Shaniv2018}. These high angular momentum states are more sensitive to pulse errors, and, therefore, robust pulse sequences are required to extend the coherence time~\cite{Genov2017,Khodjasteh2005,Souza2011,Cai2012,Kabytayev2014}.

In this work we apply rf composite pulse Ramsey sequences based on the universal rotation (UR) method with 10 rephasing pulses (UR10)~\cite{Genov2017} for precision metrology in the meta-stable $^2F_{7/2}$ manifold of a trapped $^{172}$Yb$^+$ ion. We demonstrate that the UR10 sequence is more robust than the less complex GSE method and implemented it in a trapped ion for the first time to test LLI~\cite{Dreissen2022}. We compare the two rf sequences both theoretically and experimentally in the more simple spin-$1/2$ $^2S_{1/2}$ electronic ground state of $^{172}$Yb$^{+}$ and show that the UR10 approach is 38 (13) times less sensitive to pulse duration (frequency detuning) errors than the GSE sequence. We also simulate both sequences in a 8-level system at a 1\,s Ramsey dark time. From simulation, the UR10 only introduces an error of $\epsilon_{\text{UR10}}\,=\,3\times10^{-10}$. We apply the UR10 sequence in the $^2F_{7/2}$ state and demonstrate a coherent signal of up to 2.5\,s with $10^4$ rephasing pulses. This sequence has enabled the most stringent test of Lorentz symmetry in the electron-photon sector to date, in which all the sublevels of the $^2F_{7/2}$ manifold were used, including the most sensitive $m_{J}=\pm7/2$ substates as presented in reference~\cite{Dreissen2022}. The method is also suitable for accurate measurements of quadrupole shifts to determine the quadrupole moments of meta-stable quantum states~\cite{Shaniv2016atomic}. We demonstrate this in the $^2F_{7/2}$ state and extract the quadrupole moment to be $\Theta\,=\,-0.0298(38)\,ea^{2}_{0}$. This is in agreement with the value deduced from optical clock measurement~\cite{Lange2020coherent}. With a clever choice of the magnetic field orientation, we show that an accuracy can be achieved that is competitive with the result from optical clock operation at the 10$^{-18}$ level~\cite{Lange2020coherent}.

\section{Theoretical description of the studied rf-sequences\label{sec:theory}}
It was first proposed that the principle of LLI could be tested with composite pulse Ramsey spectroscopy in reference~\cite{Shaniv2018}. We here closely follow the same theoretical description of the physical system.

\subsection{Physical system}
\begin{figure}[ht]
    \centering
    \includegraphics[width=0.5\textwidth]{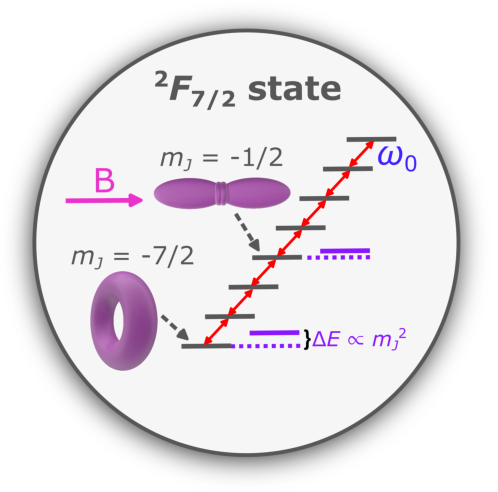}
    \caption{Zeeman substates within the $^{2}F_{7/2}$ manifold of $^{172}$Yb$^{+}$. With a defined quantization magnetic field $\textbf{B}$, the $^{2}F_{7/2}$ manifold splits into eight Zeeman sublevels spanning from $m_{J}\,=\,-7/2$ to $m_{J}\,=\,+7/2$. The $m_{J}^{2}$ dependent terms in the Hamiltonian lead to a non-linear energy shift $\Delta E$ caused by the quadrupole shift and a potential Lorentz violation.}
    \label{fig:H_exp}
\end{figure}

We consider a system with total angular momentum $J$ and magnetic sublevel $m_{J}\in\left[-J,+J\right]$, denoted by $\ket{J,m_{J}}$, interacting with an applied quantization magnetic field (B-field) $\mathbf{B} = B_z \mathbf{\hat{z}}$ via its magnetic moment $\mu_z$. The Hamiltonian describing the free evolution of the quantum system contains a linear and a quadratic part according to
\begin{equation}
     H_{\text{free}}\,=\,H_{\text{lin}} + H_{\text{quad}}\,=\,\mu_{z}\,B_{z}\,J_{z} + \kappa\,J_z^{2}\,\text{,}
     \label{equ:hamiltonianfree}
\end{equation}
where $J_{z}$ is the projection of the spin onto the quantization axis ($\mathbf{\hat{z}}$) and $\kappa$ is a parameter quantifying the magnitude of the quadratic part of the Hamiltonian. As shown in figure~\ref{fig:H_exp} for the $^{2}F_{7/2}$ manifold, the linear term induces an equally spaced energy splitting between the $m_{J}$ substates from the first-order Zeeman effect. The quadratic part describes a $m_{J}^2$-dependent energy shift $\Delta E$ due to the quadrupole shift (QS) of the state and a potential violation of the LLI denoted with the subscript LV, i.e., $\kappa\,=\,\kappa_{\text{QS}}+\kappa_{\text{LV}}$. A signal for a possible Lorentz violation is encoded in small temporal modulations of $\kappa_{\text{LV}}$ at harmonics of Earth's rotation frequency and its harmonics~\cite{Dreissen2022,Shaniv2018}.

When a driving rf field with an angular frequency of $\omega_{\text{rf}}\,=\,\omega_{0} + \delta\omega$ is applied to the system, where $\omega_{0}\,=\,(\mu_z B_z)/\hbar$ is the frequency of the linear Zeeman splitting shown in figure~\ref{fig:H_exp} and $\delta\omega$ is a small frequency detuning, the coupling Hamiltonian is given by
\begin{equation}
    H_{\text{coup}}\,=\,\Omega\,\cos(\omega_{\text{rf}}\,t + \phi)\,J_x\,\text{.}
    \label{eq:hamiltoniancoup}
\end{equation}
Here, $J_{x}$ is the generalized Pauli matrix, $\Omega$ is the multi-level Rabi frequency and $\phi$ is the phase of the rf pulse.

The total Hamiltonian of the state is the sum of the $H_{\text{free}}$ and $H_{\text{coup}}$ according to $H_{\text{tot}}\,=\,H_{\text{free}}+H_{\text{coup}}$. In the interaction picture and after applying the rotating-wave approximation, the Hamiltonian becomes
\begin{equation}
    H_{\text{tot}} = \delta\omega\,J_{z} + \kappa\,J_{z}^{2} + \Omega\,[J_{x}\,\cos(\phi) - J_{y}\,\sin(\phi)]\,\text{.}
    \label{Hamiltoniantot}
\end{equation}

In the following sections we investigate composite rf pulse sequences that suppresses the influence from $\delta\omega$ to extend the coherence time and accurately measure $\kappa$.

\subsection{Modelling Ramsey-type experiments with dynamical decoupling}
The response of a quantum system to a specific rf pulse sequence is modelled by calculating the final state $\ket{\Psi_{\mathrm{f}}}$ after applying the rf sequence to an initial state $\ket{\Psi_{\mathrm{i}}}$. This is done numerically by applying a combination of free evolution operators $T_{t}$ and rf pulse operators $U_{t,\phi}$ to $\ket{\Psi_{\mathrm{i}}}$. We denote $\tau_{0}$ as the time at which the previous state was calculated, and $\phi$ as the phase of the rf pulse.

For a specific dark time $t\,=\,t_w$, the state evolves freely as 
\begin{equation}
     \ket{\Psi(\tau_{0} + t_w)}\,=\,T_{t_w}\ket{\Psi(\tau_{0})}\,=\,\exp\left[-i\,\hbar\,t_w\,H_{\text{tot}}(\Omega\,=\,0,\phi)\right]\, \ket{\Psi(\tau_{0})}\,\text{,}
     \label{equ:freeevolu}
\end{equation}
and during a rf pulse of pulse duration $t\,=\,t_p$, the state evolves according to
\begin{equation}
     \ket{\Psi(\tau_{0} + t_p)}\,=\,U_{t_p,\phi}\ket{\Psi(\tau_{0})}\,=\,\exp\left[-i\,\hbar\,t_p\,H_{\text{tot}}(\Omega,\phi)\right]\, \ket{\Psi(\tau_{0})}\,\text{.}
     \label{equ:rfevolu}
\end{equation}

Here, we assume that changes in $\kappa$ and $\delta\omega$ are slow compared to $t_w$ and $t_p$. A superposition of $m_{J}$ substates is first created with a $\pi/2$-pulse of duration $t_{p}\,=\,\pi/(2\Omega)$ by applying the operator $U_{\pi/2,0}$. Then the state evolves freely  by applying the operator $T_{t_{w}}$, after which a $\pi$-pulse of duration $t_{p}\,=\,\pi/\Omega$ with a specific phase $\phi_{i}$ is applied using the operator $U_{\pi,\phi_{i}}$ and another dark time of $t_{w}$ follows. The phases of the $n$ rf pulses in the specific sequences investigated in this work are given in table~\ref{tab:phases}. The composite rf pulse sequences consist of $n$ repetitions of the combination [$T_{t_{w}}$]\,-\,[$U_{\pi,\phi_{i}}$]\,-\,[$T_{t_{w}}$]. After the composite rf pulse sequence, a final $\pi/2$-pulse with a phase of $\pi$ with respect to the first $\pi/2$-pulse is applied with the operator $U_{\pi/2,\pi}$ to measure the acquired phase during the dark time via the fraction of the population that is retrieved back into the initial $m_{J}$ state. Thus we can write the final state as
\begin{equation}
    \ket{\Psi_{\mathrm{f}}}\,=\,U_{\pi/2,\, \pi}\cdot\prod^{n}_{i\,=\,1} \left( T_{t_w}\cdot U_{\pi,\, \phi_i}\cdot T_{t_w} \right)\cdot U_{\pi/2,\, 0} \ket{\Psi_{\mathrm{i}}}\,\text{.}
    \label{equ:finalstateexpressshort}
\end{equation}

\begin{table}[ht]
    \centering
    \begin{tabular}{llc}
      \hline\hline 
      Sequence  & Phases $\phi_{i}$ & Repetition $n$ \\ \hline
      GSE    & (1, -1) $\pi$/2 & 2\\
      UR10   &  (0, 4, 2, 4, 0, 0, 4, 2, 4, 0) $\pi$/5 & 10\\ \hline\hline
    \end{tabular}
    \caption{Phases $\phi_i$ of the $\pi$-pulses for the generalized spin-echo (GSE) sequence and universal rotation sequence with 10 rephasing pulses (UR10).}
    \label{tab:phases}
\end{table}

In order to extend the Ramsey dark-time to several seconds, the modulation sequence is repeated $N_{\text{rep}}$ times. The final state is thus given by
\begin{equation}
    \ket{\Psi_{\mathrm{f}}}\,=\,U_{\text{tot}}\ket{\Psi_{\mathrm{i}}}\,=\,U_{\pi/2,\, \pi}\cdot\prod^{N_{\text{rep}}}_{N\,=\,1}\prod^{n}_{i\,=\,1} \left( T_{t_w}\cdot U_{\pi,\, \phi_i}\cdot T_{t_w} \right)_{N}\cdot U_{\pi/2,\, 0} \ket{\Psi_{\mathrm{i}}}\,\text{.}
    \label{equ:finalstateexpress}
\end{equation}

The total Ramsey dark time $T_{\text{D}}\,=\,2n\cdot N_{\text{rep}}\cdot t_{w}$ is given by free evolution time, i.e., the sum of all the $t_{w}$. The pulse sequence cancels dephasing from the linear contribution in the Hamiltonian, proportional to $\delta\omega$, and the phase that is acquired during the dark time is only dependent on energy shifts induced by the quadratic part proportional to $\kappa$. The value of $t_{w}$ that allows for successful rephasing of the state vector is determined by the ambient magnetic field noise in the experimental environment. In our case, $t_{w}$ was limited to $t_{w}\,\lesssim\,200$\,µs. 

\subsection{Simulation of robustness against rf pulse imperfections}
In order to use this method for a sensitive test of LLI in the $^{2}F_{7/2}$ state of Yb$^{+}$ and to extract the quadrupole moment of this state, a Ramsey dark time of on the order of one second should be reached, in which case several thousands of rephasing pulses are applied. Small imperfections in these rf pulses, such as a detuning or a pulse duration error, accumulate quickly and induce dephasing of the atomic state before the required dark time is reached. Therefore, we characterize the robustness of the studied sequences against these types of pulse errors. 

\begin{figure}[ht]
    \centering
    \includegraphics[width=0.8\textwidth]{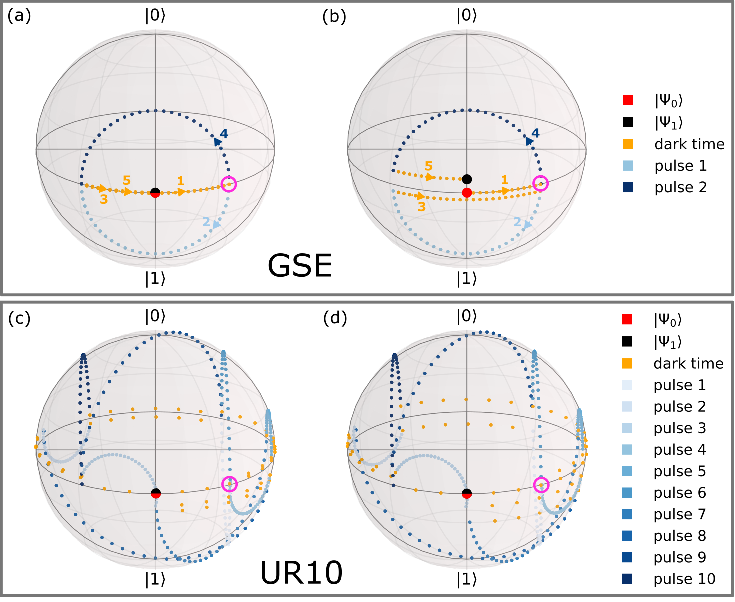}
    \caption{Bloch sphere representation of the GSE and the UR10 sequences with one repetition $N_{\text{rep}}\,=\,1$. The rf detuning $\delta\omega$ is scaled to the rf Rabi frequency $\Omega$ as $\eta_{\omega}\,=\,\delta\omega/\Omega$ and pulse duration error $\delta t$ is scaled to the $\pi$-pulse duration $t_{\pi}$ as $\eta_{t}\,=\,\delta t/t_{\pi}$. The states $\ket{\Psi_{0}}$ and $\ket{\Psi_{1}}$ are denoted in red and black, respectively. The purple circle indicates the start of the first $\pi$-pulse. Figure (a) and (c) show the state evolution of the GSE and the UR10 sequences, respectively, with $\eta_{\omega}\,=\,- 0.03$ and $\eta_{t}\,=\,0$. Figure (b) and (d) show the state evolution of the GSE and UR10 sequences, respectively, with $\eta_{\omega,t}\,=\,- 0.03$. For figure (a) and (b), the state starts off at the equator at the red dot and after a combination of three periods of free evolution and two $\pi$-pulses with specific phases, lands at the location shown with the black dot. The sequence introduces an error of $\epsilon_{\text{GSE}}\,=\,3\times10^{-3}$. For (c) and (d), the state starts off at the equator at the red dot and after a combination of eleven periods of free evolution and ten $\pi$-pulses with specific phases. A simulation with $t_{w}\,=\,100$\,µs shows that the sequence introduces an error of $\epsilon_{\text{UR10}}\,=\,2\times10^{-16}$.}
    \label{fig:bs}
\end{figure}

To describe the influence of pulse errors on the rf sequence in an intuitive manner, the state evolution is shown on the Bloch sphere, see figure~\ref{fig:bs}. Here, $\ket{\Psi_{0}}\,=\,U_{\pi/2,\, 0} \ket{\Psi_{\mathrm{i}}}$, is shown in red, and $\ket{\Psi_{1}}\,=\,\prod^{n}_{i\,=\,1} \left( T_{t_w}\cdot U_{\pi,\, \phi_i}\cdot T_{t_w} \right)\ket{\Psi_{0}}$, is shown in black. The state vector is considered to be rephased successfully if the position on the Bloch sphere is the same for $\ket{\Psi_{0}}$ and $\ket{\Psi_{1}}$. A rf detuning $\delta\omega$ is scaled to the rf Rabi frequency $\Omega$ as $\eta_{\omega}\,=\,\delta\omega/\Omega$ and a pulse duration error $\delta t$ is scaled to the $\pi$-pulse duration $t_{\pi}$ as $\eta_{t}\,=\,\delta t/t_{\pi}$. The purple circle indicates the start of the first $\pi$-pulse. Figure~\ref{fig:bs}\,(a) shows the state evolution with the GSE sequence for $\eta_{\omega}\,=\,- 0.03$ and $\eta_{t}\,=\,0$, while in figure~\ref{fig:bs}\,(b) both pulse errors are set to $\eta_{\omega,t}\,=\,- 0.03$. Figure~\ref{fig:bs}\,(c) and (d) shows the state evolution with the UR10 sequence for the same values of $\eta_{\omega,t}$ as in (a) and (b), respectively. At $N_\text{rep}\,=\,1$, $t_{w}\,=\,100$\,µs and with a pulse error of $\eta_{\omega,t}\,=\,- 0.03$, the GSE sequence introduces an error of $\epsilon_{\text{GSE}}\,=\,1-|\langle\Psi_{1}|\Psi_{0}\rangle|^{2}\,=\,3\times10^{-3}$. The error would accumulate as $N_\text{rep}$ increases and the spin can no longer be rephased. For the UR10 sequence, the spin rephasing only introduces an error of $\epsilon_{\text{UR10}}\,=\,2\times10^{-16}$. Therefore, this sequence can be used to suppress first order dephasing effects while having detuning and pulse duration errors.

To investigate the tolerance of the rf sequences to these types of pulse errors quantitatively, the fidelity can be calculated according to
\begin{equation}
    \mathcal{F}(\eta_{\omega}, \eta_{t}) = |\bra{\Psi_{\text{ideal}}} U_{\text{tot}}(\eta_{\omega}, \eta_{t}) \ket{\Psi_{\mathrm{i}}}|^2\,\text{,}
    \label{eq:fidelity}
\end{equation} 
where the frequency detuning $\eta_{\omega}$ and the pulse duration error $\eta_{t}$ are introduced in each evolution operator and $\ket{\Psi_{\text{ideal}}} = U_{\text{tot}}(\eta_{\omega}\,=\,0, \eta_{t}\,=\,0) \ket{\Psi_{\mathrm{i}}}$ is the ideal outcome of an experiment. For the numerical implementation of equation~\ref{equ:finalstateexpress} in Python, the QuTip package~\cite{Johansson2012,Johansson2013} was used to provide the generalized Pauli matrices $J_{x/y/z}$ for arbitrary states $J$. In the following sections, the fidelity will be simulated for different $\eta_{\omega,t}$ and compared with experimental results to investigate the robustness of the composite rf pulse sequences.

\section{Investigation of the robustness of rf pulse sequences\label{sec:rfseqinvest}}
%\subsection{General measurement scheme}
For accurate spectroscopic measurements using the rf sequence in the $^{2}F_{7/2}$ manifold, coherent population transfer to and from the $^2F_{7/2}$ manifold via the highly forbidden electric octupole (E3) transition from the electronic ground state $^{2}S_{1/2}$ is required. Excitation of the ion via the E3 transition is slow; at a typical Rabi frequency of $\Omega\,=\,2\pi\times$10\,Hz, full population transfer takes about $t_{\pi}\,=\,\pi/\Omega\,=\,50$\,ms~\cite{Fuerst2020}. To gather enough statistics, each measurement is repeated 50 times, during which significant drifts of the rf power and the quantization B-field occur. Therefore, it is impractical to map out the stability of the rf sequences experimentally in the $^{2}F_{7/2}$ manifold. Instead, we first investigated the sequences in the spin-$1/2$ $^{2}S_{1/2}$ electronic ground state of the $^{172}$Yb$^{+}$ ion, where state preparation is simple and no ultra-stable laser is required. We then implement and test the sequence that meets the requirements in the $^{2}F_{7/2}$ manifold. 

For simplicity, from now on, the notations $\ket{S,m_{J}}$, $\ket{D,m_{J}}$ and $\ket{F,m_{J}}$ will be used for the Zeeman sublevels of the $^{2}S_{1/2}$, $^{2}D_{5/2}$ and $^{2}F_{7/2}$ manifolds, respectively.

\subsection{Experimental setup}\label{subsec:experimental setup}
\begin{figure}[ht]
    \centering
    \includegraphics[width=0.8\textwidth]{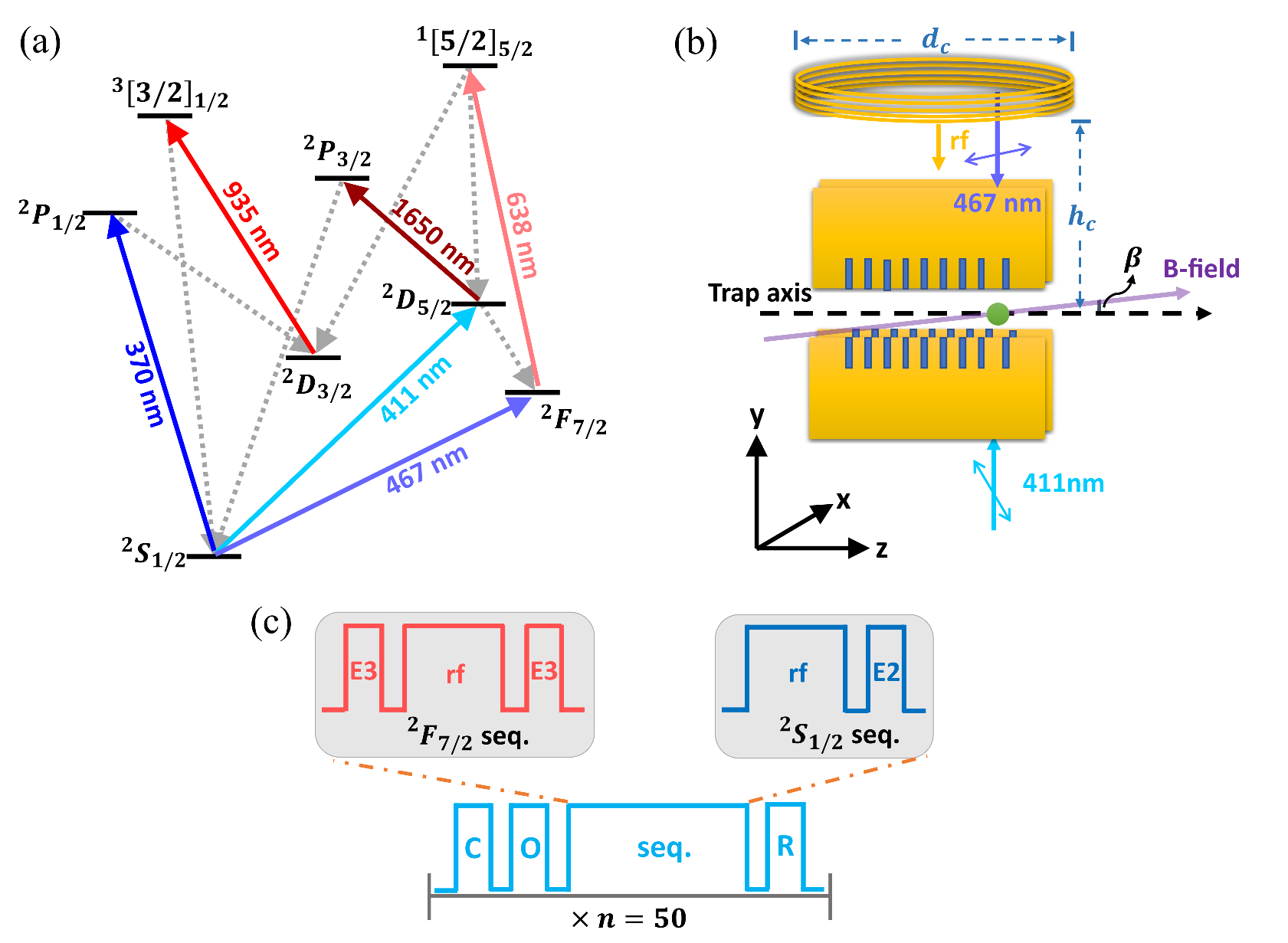}
    \caption{(a) Reduced energy level diagram of $^{172}$Yb$^{+}$. Doppler cooling and repumping are carried out on the transitions near 370\,nm and 935\,nm, respectively. Optical pumping into either one of the $\ket{S,\pm1/2}$ states is done via a $\sigma^{\pm}$-polarized 370\,nm laser beam parallel to the magnetic field (B-field). Repumpers near 1650\,nm and 638\,nm are used to bring the population back to the ground state after excitation to the $^{2}D_{5/2}$ or $^{2}F_{7/2}$ state on the transitions near 411 or 467\,nm, respectively. (b) Laser and quantization B-field orientation. The B-field lies in the $xz$ plane at an angle of $\beta\,=\,26.8(4.0)$\,$^{\circ}$ with the trap axis $\mathbf{\hat{z}}$. The 467\,nm (411\,nm) beam points in the $-\mathbf{\hat{y}}$ ($+\mathbf{\hat{y}}$) direction with a polarization parallel (perpendicular) to the B-field. The resonant circuit including an antenna coil used for rf spectroscopy has a diameter of $d_{c}\,=\,$4.5\,cm and is mounted $h_{c}\,=\,$5.5\,cm above the ion. (c) Reduced experimental sequence for rf spectroscopy in the $^{2}S_{1/2}$ manifold (``$^{2}S_{1/2}$ seq.") and in the $^{2}F_{7/2}$ manifold (``$^{2}F_{7/2}$ seq."). ``C" denotes Doppler-cooling via the 370\,nm transition, ``O" denotes optical pumping, and ``rf" is the rf sequence. ``E2" and ``E3" represent the excitation via the electric quadrupole and electric octupole transitions, respectively. ``R" is the repumping via transitions near 1650\,nm and 638\,nm.}
    \label{fig:levelscheme}
\end{figure}

We carried out the experimental investigation of the rf sequences with an $^{172}$Yb$^{+}$ ion that is trapped in a segmented rf Paul trap, for details of the trap see reference~\cite{Keller2019}. Figure~\ref{fig:levelscheme}\,(a) shows a reduced energy level diagram with the relevant transitions. The ion is cooled to the Doppler temperature of $T\,\approx$\ 0.5\,mK via the dipole allowed $^{2}S_{1/2}\,\rightarrow\,^{2}P_{1/2}$ transition near 370\,nm. Spontaneous decay occurs from the $^{2}P_{1/2}$ state to the $^{2}D_{3/2}$ state. Via the $^{2}D_{3/2}\,\rightarrow\,^{3}[3/2]_{1/2}$ transition near 935\,nm, the population is pumped back to the cooling cycle. The same cycle is used for fluorescence detection of the internal state of the ion. A circularly polarized $\sigma^{\pm}$ laser beam on resonance with the 370\,nm transition, propagating parallel to the quantization B-field, is used for selective optical pumping into either one of the $\ket{S,\pm 1/2}$ states. The electric quadrupole (E2) transition near 411\,nm can be used to excite the ion to the $^{2}D_{5/2}$ state. The 411\,nm beam of $P_{411}\,\approx$\ 0.4\,mW is focused down to a waist of $w_{411}\,\approx\ $83\,µm at the position of the ion to reach a Rabi frequency of $\Omega_{\text{E2}}\,\approx\,2\pi\times 31.3$\,kHz. From this state, there is a probability that the ion decays to the $^{2}F_{7/2}$ state, which has a very long lifetime of approximately 1.6 years~\cite{Lange2021}. Therefore, the $^{2}F_{7/2}\,\rightarrow\,^{1}[5/2]_{5/2}$ and $^{2}D_{5/2}\,\rightarrow\,^{2}P_{3/2}$ transitions near 638\,nm and 1650\,nm are used as repumpers to bring the population back the electronic ground state. The E3 transition near 467\,nm is used to coherently transfer the population to and from the $^{2}F_{7/2}$ state. 

For the orientation of the quantization B-field and the probe beams near 411\,nm and 467\,nm, see figure~\ref{fig:levelscheme}\,(b). The B-field lies in the $xz$ plane at an angle of $\beta\,=\,$26.8(4.0)\,$^{\circ}$ with the trap axis $\mathbf{\hat{z}}$. The magnitude of the field can be tuned between $B\,=\,$50\,-\,250\,µT. The 467\,nm (411\,nm) laser beam points in the $-\mathbf{\hat{y}}$ ($+\mathbf{\hat{y}}$) direction and is polarized parallel (perpendicular) to the B-field. 

For interrogation of the E3 transition, a beam of 6\,mW is focused down to a waist of $(w_{x},w_{y})\,\approx\,(26,38)$\,µm at the ion to reach a Rabi frequency of $\Omega_{\text{E3}}\,\approx\,2\pi\times$10\,Hz. At this Rabi frequency, the full-width-half-maximum (FWHM) linewidth of the $|S,\pm1/2\rangle\,\rightarrow\,|F,\pm1/2\rangle$ E3 transition is $\nu_{\text{FWHM}}\,\approx\,$20\,Hz, while this transition has a Zeeman sensitivity of $\delta\nu(\Delta m\,=\,0)\,=\,6$\,GHz/T. Therefore, the B-field noise needs to be reduced to the level of $\delta B\,<\,1$\,nT to achieve coherent E3 excitation within $\tau_{\pi}\,=\,$50\,-\,100\,ms. We reach the required stability at frequencies of up to 550\,Hz by applying active feedback on the B-field. A magnetic field sensor~\footnote{SENSYS Magnetometer \& Survey Solutions, SENSYS FGM3D} mounted $h_{s}\,\approx\ $8.5\,cm above the ion is used in the feedback loop. A resonant circuit including an antenna coil is used for rf spectroscopy. It has a diameter of $d_{c}\,=\ $4.5\,cm and is mounted $h_{c}\,=\ $5.5\,cm above the ion to produce a field along the $-\mathbf{\hat{y}}$ direction (see figure~\ref{fig:levelscheme}(b)) at a resonance frequency of $\nu_{c}\,=\ $3.5147(7)\,MHz with a temperature sensitivity of $d\nu/dT\,=\,$7.3\,kHz/$^{\circ}$C. The pulses are generated by a AD9910 direct digital synthesizer in the Sinara hardware~\footnote{https://m-labs.hk/experiment-control/sinara-core/}. The phases of the pulses have to be a rational number in units of $\pi$, which facilitates the implementation of the UR10 sequence.
%For detailed configuration of the coil, see~\ref{ap:rf_coil}.

\subsection{RF resonance frequency and Rabi frequency\label{subsec:rfscan}}
To implement the rf sequences, we first measure the resonance frequency of the Zeeman splitting and the rf Rabi frequency in both the $^2S_{1/2}$ and the $^2F_{7/2}$ manifolds. The sequences used for these measurements are shown in figure~\ref{fig:levelscheme}\,(c). In both cases, the ion is first cooled to Doppler temperature and then the population is optically pumped to the $\ket{S,-1/2}$ state. In the case of the $^{2}S_{1/2}$ sequence, a single rf pulse is then applied, followed by excitation on the $|S,-1/2\rangle\,\rightarrow\,|D,-5/2\rangle$ E2 transition to shelve the population to the $^{2}D_{5/2}$ state for state selective fluorescence detection. For the $^{2}F_{7/2}$ sequence, a $|S,-1/2\rangle\,\rightarrow\,|F,-1/2\rangle$ E3 excitation pulse is applied before and after the rf pulse. In both cases, fluorescence detection is used to determine the internal state of the ion and a repumping sequence is applied to prepare the ion for the next measurement run. 

\begin{figure}[ht]
    \centering
    \includegraphics[width=0.8\textwidth]{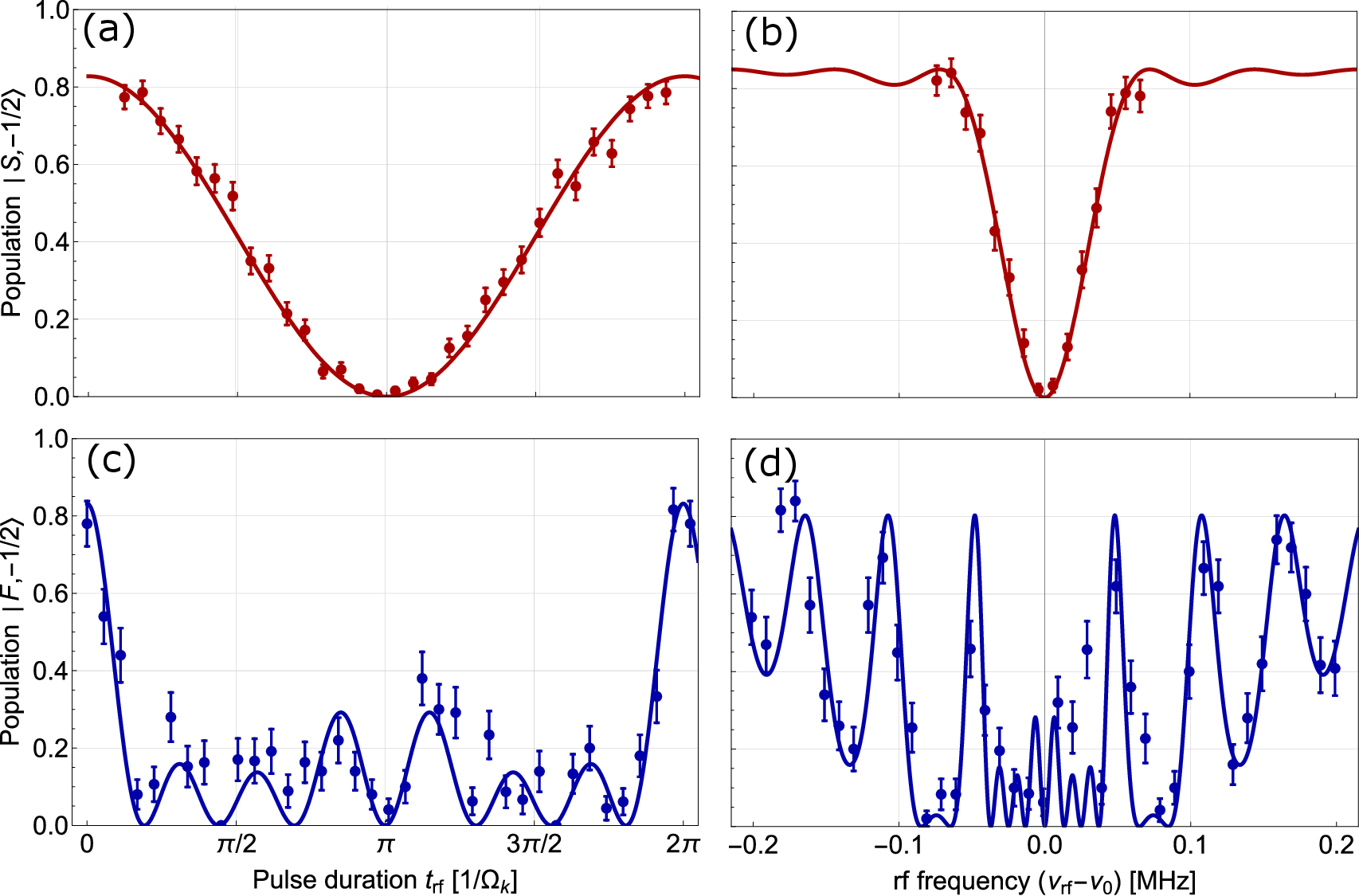}
    \caption{Example of measurements to determine (a,c) the Rabi frequency and (b,d) the rf center frequency in both the $^{2}S_{1/2}$ and $^{2}F_{7/2}$ manifold. For both measurements in the $^{2}S_{1/2}$ and $^{2}F_{7/2}$ manifolds, the ion is prepared in the $m_{J}\,=\,-1/2$ state. The population that remains in the initial state after applying a single rf pulse is measured, from which the Rabi frequency and the resonance frequency are extracted. (a) The two-level Rabi frequency in the $^{2}S_{1/2}$ state is $\Omega_{S}\,=\,2\pi\times 60.5(4)$\,kHz. (b) The measured rf center frequency in the $^{2}S_{1/2}$ state is $\nu_{0}\,=\,3.5942(7)$\,MHz. (c) The multi-level Rabi frequency in the $^{2}F_{7/2}$ state is $\Omega_{F}\,=\,2\pi\times 28.1(1)$\,kHz. (d) The measured rf center frequency in the $^{2}F_{7/2}$ state is $\nu_{0}\,=\,3.5510(6)$\,MHz. The fluctuations in (c,d) comes from slow drifts of the E3 transition frequency during the measurement.}
    \label{fig:rfscan}
\end{figure}

Figure~\ref{fig:rfscan}\,(a) and (b) show typical measurements in the $^{2}S_{1/2}$ manifold with each measurement point averaged over 100 repetitions and figure~\ref{fig:rfscan}\,(c) and (d) show those in the $^{2}F_{7/2}$ manifold with an averaging of 50 repetitions. The y-axis denotes the population that is still left in the initial state after applying a single rf pulse. Either the pulse duration (figure~\ref{fig:rfscan}\,(a) and (c)) or the applied rf frequency (figure~\ref{fig:rfscan}\,(b) and (d)) is varied. The pulse duration is scaled to $1/\Omega_{k}$, where $\Omega_{k}$ is the Rabi frequency in either the $^{2}S_{1/2}$ manifold ($k\,=\,S$) or the $^{2}F_{7/2}$ manifold ($k\,=\,F$). The population is transferred fully from the $m_{J}\,=\,-1/2$ state to the $m_{J}\,=\,+1/2$ state at $\nu_{\text{rf}}\,=\,\nu_{0}$ and $t_{\text{rf}}\,=\,\pi/\Omega_{k}$. From the measurements shown in figure~\ref{fig:rfscan}\,(a) and (b), a Rabi frequency of $\Omega_{S}\,=\,2\pi\times 60.5(4)$\,kHz and a resonance frequency of $\nu_{0}\,=\,3.5942(7)$\,MHz are extracted at a quantization field of $B\,\approx\,128$\,µT. From the measurements conducted in the $^{2}F_{7/2}$ manifold, a Rabi frequency and resonance frequency of $\Omega_{F}\,=\,2\pi\times 28.1(1)$\,kHz and $\nu_{0}\,=\,3.5510(6)$\,MHz  are extracted, respectively, at a quantization field of $B\,\approx\,222$\,µT. The finite excitation probability of about 85\,\% on the E2 transition originates from the $7\,$ms lifetime of the $^{2}D_{5/2}$ state, thermal decoherence from average phonon occupation number of $\bar{n}\,=\,15$ and imperfect polarization and pulse duration of the 411\,nm beam. The fluctuations in figure~\ref{fig:rfscan}\,(c) and (d) are caused by slow drifts of the E3 transition frequency during the measurement.

In figure~\ref{fig:rfscan}\,(b) and (d), the FWHM linewidth of the rf center frequency in the $^{2}S_{1/2}$ manifold is $\nu_{\text{FWHM},S}\,\approx\,60$\,kHz at $\Omega_{S}\,=\,2\pi\times 60.5(4)$\,kHz, while in the $^{2}F_{7/2}$ manifold it is $\nu_{\text{FWHM},F}\,\approx\,10$\,kHz at $\Omega_{F}\,=\,2\pi\times 28.1(1)$\,kHz. The ratio of these quantities, $\nu_{\text{FWHM},S}/\Omega_{S}\,\approx\,2.8\times\nu_{\text{FWHM},F}/\Omega_{F}$, shows that the relative width of the $^{2}F_{7/2}$ resonance is 2.8 times smaller compared to that in the $^{2}S_{1/2}$ manifold at similar Rabi frequencies. Hence, the state in the $^{2}F_{7/2}$ manifold is more sensitive to fluctuations in rf power and magnetic field environment causing pulse errors. 

\subsection{Comparison between simulations and experiments in the $^2S_{1/2}$ state\label{sec: 2level_verify}}
We now investigate the fidelity of a rf pulse sequence as a function of the relative pulse errors, $\eta_{\omega}$ and $\eta_{t}$ in a spin-$1/2$ quantum system. For comparison, we perform measurements in the $^{2}S_{1/2}$ electronic ground state of $^{172}$Yb$^{+}$ to characterize various regions of the stability diagram experimentally.

The sequence used for the measurements (see ``$^{2}S_{1/2}$ seq." in figure~\ref{fig:levelscheme}\,(c)) is similar to the one described in section~\ref{subsec:rfscan}, but now the single rf pulse is replaced with the full composite rf pulse sequence and the E2 shelving pulse addresses the $|S,+1/2\rangle\,\rightarrow\,|D,+5/2\rangle$ transition. The experimental fidelity of the rf sequences is defined in the following way. If the ion appears bright with fluorescence detection, shelving via the E2 transition was not successful and the population is in the $|S,-1/2\rangle$ state as intended, in which case $\mathcal{F}$\,=\,1 (see equation~\ref{eq:fidelity}). In the opposite case, the ion appears dark, meaning that the population was transferred to the $\ket{S,+1/2}$ state and then shelved to the $^2D_{5/2}$ manifold, i.e., $\mathcal{F}$\,=\,0. Note that the final Ramsey $\pi/2$-pulse has a phase of $\phi\,=\,\pi$, which brings the population back to the initial $|S,-1/2\rangle$ state at the end of the rf sequence. Each measurement point is averaged over 50 repetitions. The finite excitation probability of about 85\,\% on the E2 transition originates from the $7\,$ms lifetime of the $^{2}D_{5/2}$ state, thermal decoherence from average phonon occupation number of $\bar{n}\,=\,15$ and imperfect polarization and pulse duration of the 411\,nm beam. To compare the experimental results with the simulated fidelity, the background of 15\,\% is subtracted and the data is re-normalized.

For a long coherence time within the $^{2}F_{7/2}$ manifold, the fidelity of the rf sequence is required to be high. The quantitative minimum requirements we set on the robustness of the sequences are given by experimental parameters. For Ramsey spectroscopy in the $^{2}F_{7/2}$ manifold, the excitation probability on the E3 transition and the fidelity of the rf sequence determines the contrast and, as a result, the measurement sensitivity. Since the E3 excitation probability is around 0.85, a minimum requirement for the fidelity of $\mathcal{F}\,\geqslant\,0.85$ at $T_{D}\,\approx\,1$\,s is set. Furthermore, from the measurements shown in figure~\ref{fig:rfscan}\,(c) and (d), the rf resonance frequency can only be determined with an accuracy of $\sigma_{\eta_{\omega}}\,=\,\pm 0.021$ and the pulse duration with an accuracy of $\sigma_{\eta_{t}}\,=\,\pm 0.004$. Therefore, we require the fidelity for the rf sequences to be $\mathcal{F}\,\geqslant\,0.85$ in the rf pulse error range of $-0.021\,<\,\eta_{\omega}\,<\,0.021$ and $-0.004\,<\,\eta_{t}\,<\,0.004$ in our experiment.

\begin{figure}[ht]
    \centering
    \includegraphics[width=0.8\textwidth]{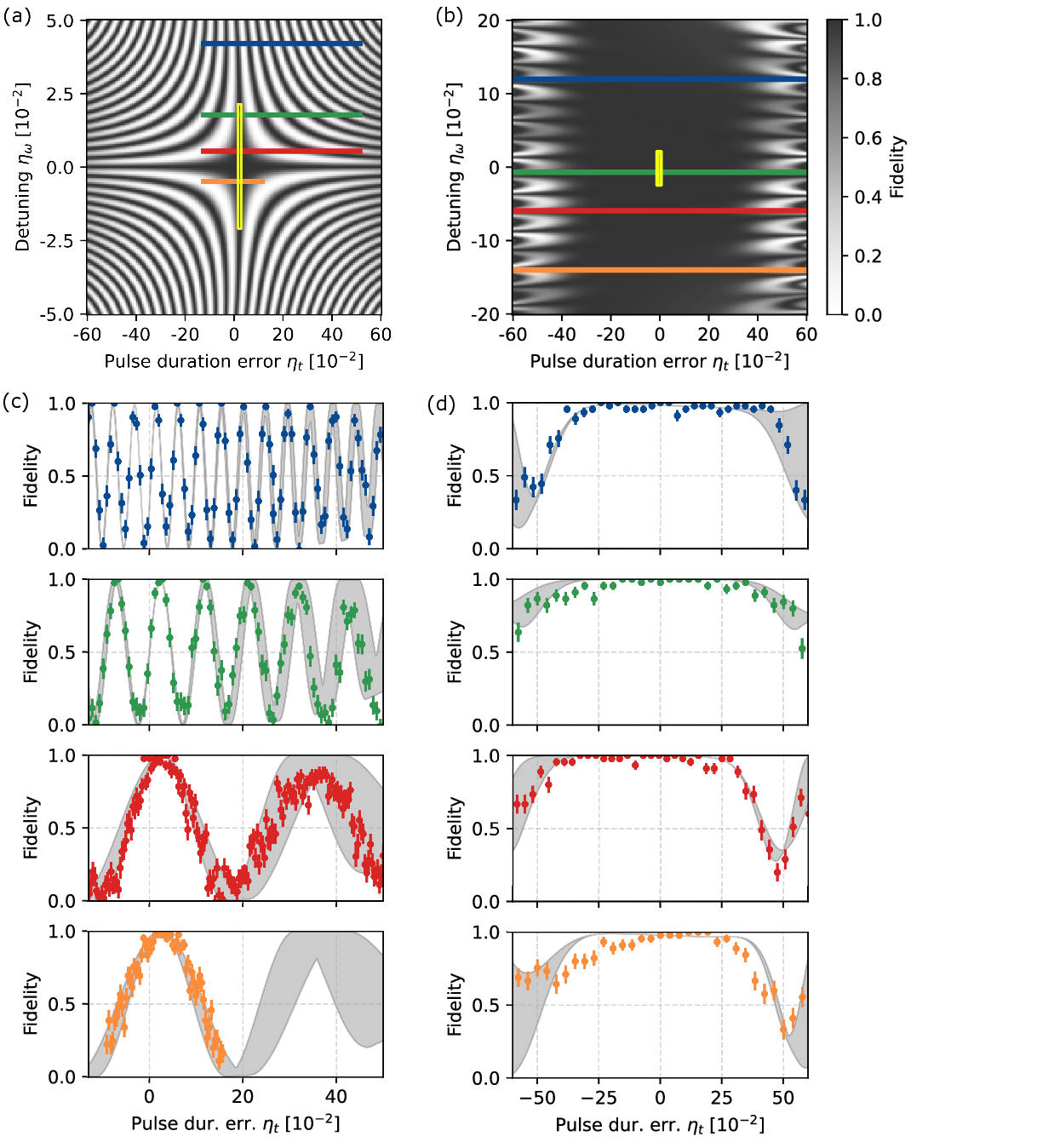}
    \caption{Simulation and experimental verification of the GSE and UR10 sequences in a spin-$1/2$ system at a Ramsey dark time of $T_{\text{D}}\,=\,$10\,ms. Figure (a) shows the stability diagram for pulse errors $\eta_{\omega}$ and $\eta_{t}$ for the GSE sequence and figure (b) shows the UR10 sequence. The grey scale indicates the fidelity from 0 to 1. The colored lines show the region where the sequences are investigated experimentally in the $^{2}S_{1/2}$ electronic ground state of $^{172}$Yb$^{+}$. The yellow box indicates the threshold that was set in our experiment. Figure (c) and (d) show the experimental results of the GSE and the UR10 sequence, respectively. The colors of the data points correspond to the region indicated in (a) and (b), and the grey solid areas show the simulated results with a typical B-field fluctuation of $\delta B\,=\,$1\,-\,2\,nT within a measurement time of 4\,-\,5\,min. Each measurement point is averaged over 50 repetitions. From the simulation, the UR10 sequence tolerates 38 and 13 times larger values of $\eta_{t}$ and $\eta_{\omega}$, respectively, compared to the GSE sequence.}
    \label{fig:map_all}
\end{figure}

To verify that the simulations of the GSE and the UR10 agree with our experiment, we first show the simulated fidelity of the rf sequences in the two-level $^2S_{1/2}$ system at a Ramsey dark time of $T_{\text{D}}\,=\,$10\,ms in the stability diagrams of figure~\ref{fig:map_all}\,(a) and (b). Here we simulated fixed pulse errors, i.e., errors that are the same for every $\pi$-pulse in a sequence throughout an experiment. The grey scale indicates the fidelity from 0 to 1. In these sequences, the wait time was set to $t_{\text{w}}\,=\,100$\,µs and 50 rephasing pulses were applied. 

\subsubsection{GSE sequence\label{subsubsec:GSEandUR10}}
The stability diagram of the GSE sequence is shown in figure~\ref{fig:map_all}\,(a). The colored lines in the figure indicate the range in which the sequence is investigated experimentally. The yellow box indicates the error range that was set in our experiment. The experimental results are shown in figure~\ref{fig:map_all}\,(c), where the data points are given in the same colors as the ranges indicated in figure~\ref{fig:map_all}\,(a) and the grey shaded areas are the corresponding simulated results, where a typical B-field fluctuation of $\delta B\,=\,$1\,-\,2\,nT within a measurement time of 4\,-\,5\,min was taken into account. Overall, the data agrees well with the simulations. During the measurements, the magnetic field was set to $B\,\approx\,$122\,µT, at which the rf center frequency was $\nu_{0}\,=\,3.423(1)$\,MHz and the Rabi frequency was $\Omega_{S}\,=\,2\pi\times 40.0(2)$\,kHz, resulting in a $\pi$-pulse duration of $t_{\pi}\,=\,12.5(1)$\,µs. Note that these values differ from the ones given in section~\ref{subsec:rfscan} because they were taken at different B-fields.

The simulation shows a clear fringe pattern. Within the center fringe, at a fixed value of $|\eta_{\omega}|\,=\,0.021$ corresponding to the experimental measurement uncertainty, a fidelity of $\mathcal{F}\,\geqslant\,$0.85 is reached for $|\eta_t|\,<\,0.01$. Similarly, a fidelity of $\mathcal{F}\,\geqslant\,$0.85 is reached for $|\eta_{\omega}|\,<\,0.03$ for a fixed pulse duration error $|\eta_t|\,=\,0.004$. Therefore, the GSE sequence just meets the requirements set at the beginning of section~\ref{sec: 2level_verify} in this case. However, the sequence does not allow for additional drifts of the magnetic field or the rf power. Furthermore, the simulations and measurements are done with a dark time of only 10 ms. For longer dark times, as required in the test of LLI, the acceptable stability range becomes even smaller and will be shown in section~\ref{subsec:8levelsimu}. 

\subsubsection{UR10 sequence\label{subsubsec:UR10}}
The stability diagram of the UR10 sequence is shown in figure~\ref{fig:map_all}\,(b). The experimental data (see figure~\ref{fig:map_all}\,(d)) and the simulated curves agree well also for the UR10 sequence. During this measurement, the magnetic field was $B\,\approx\,128$\,µT, at which the center frequency and the Rabi frequency were measured to be $\nu_{0}\,=\,3.5942(7)$\,MHz and $\Omega_{S}\,=\,2\pi\times38.40(6)$\,kHz, respectively, corresponding to a $\pi$-pulse duration of $t_{\pi}\,=\,13.0(2)$\,µs.

The stability diagram is much more homogeneous than that of the GSE sequence. The simulation shows that the fidelity $\mathcal{F}\,\geqslant\,$0.85 for $|\eta_{t}|\,<\,0.38$ at a fixed $|\eta_{\omega}|\,=\,0.021$. The allowed frequency detuning error is $|\eta_{\omega}|\,<\,0.4$ for a fixed $|\eta_t|\,=\,0.004$. The UR10 sequence easily meets the experimental requirements of $\mathcal{F}\,\geqslant\,$0.85 at $|\eta_{\omega}|\,<\,0.021$ and $|\eta_t|\,<\,0.004$ and, therefore, allows for additional drifts of both the magnetic field and the rf power. The UR10 sequence tolerates 38 and 13 times larger errors of $\eta_{t}$ and $\eta_{\omega}$, respectively, compared to the GSE sequence. The UR10 sequence, therefore, enables robust precision rf spectroscopy even in noisy lab environment in a spin-$1/2$ system.

\subsection{Simulation of the rf pulse sequences in an eight-level system}\label{subsec:8levelsimu}
\begin{figure}[ht]
    \centering
    \includegraphics[width=0.8\textwidth]{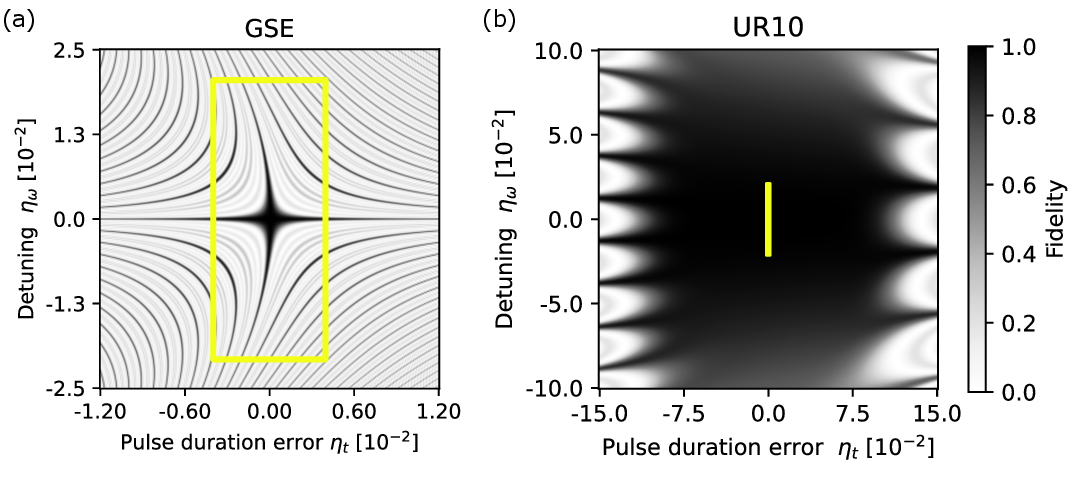}
    \caption{Simulation of the fidelity of the GSE and UR10 sequences in the eight-level system with 1\,s of Ramsey dark time. The grey scale indicates fidelity from 0 to 1. The stability diagram for (a) the GSE and (b) the UR10 sequence is obtained by varying $\eta_{\omega}$ and $\eta_{t}$. The yellow box represents the threshold of $|\eta_{\omega}|\,<\,0.021$ and $|\eta_t|\,<\,0.004$ for which we require $\mathcal{F}\,\geqslant\,0.85$. For the GSE sequence, the requirement is clearly not met, while the UR10 sequence stays above 0.85 for a much larger range of $|\eta_{\omega}|\,<\,0.06$ and $|\eta_{t}|\,<\,0.075$. Within the yellow box, the maximum error that the UR10 sequence introduces is $\epsilon_{\text{UR10}}\,=\,3\times10^{-10}$.}
    \label{fig:DD_8lvl_comp}
\end{figure}

To characterize the robustness of the GSE and UR10 sequences for high-precision Ramsey spectroscopy in the $^{2}F_{7/2}$ manifold, we simulate the stability diagram in the eight-level system at $T_{\text{D}}\,=\,$1\,s. The results for the GSE and UR10 sequence are shown in figure~\ref{fig:DD_8lvl_comp}\,(a) and (b), respectively. In both simulations, the wait time was set to $t_{w}\,=\,100$\,µs and to reach the required Ramsey dark time, a total of 5000 rephasing pulses were applied. In this case, $N_{\text{rep}}\,=\,2500$ and $N_{\text{rep}}\,=\,500$ for GSE and UR10 sequences, respectively.

The stability diagram of the GSE sequence is simulated for a range of $-0.025\,<\,\eta_{\omega}\,<\,0.025$ and $-0.012\,<\,\eta_{t}\,<\,0.012$. Even in this small error range, a narrowly spaced fringe pattern is observed. The stability diagram of the UR10 sequence is simulated for a much larger error range of $-0.10\,<\,\eta_{\omega}\,<\,0.10$ and $-0.15\,<\,\eta_{t}\,<\,0.15$ and looks more homogeneous. Given our experimental uncertainties of $|\eta_{\omega}|<0.021$ and $|\eta_{t}|<0.004$, the GSE sequence is clearly not robust enough for our application. This is further illustrated with the yellow boxes shown in the figure, which indicate the bounds of our experimental uncertainties of $\eta_{\omega}$ and $\eta_{t}$. The UR10 sequence tolerates much larger errors of $|\eta_{\omega}|\,<\,0.06$ and $|\eta_{t}|\,<\,0.075$ to maintain a $\mathcal{F}\,\geqslant\,$0.85. Within the experimental uncertainties, the error that the UR10 sequence introduces is only $\epsilon_{\text{UR10}}\,=\,1-\mathcal{F}\,=\,3\times10^{-10}$. Hence, the UR10 sequence can be used for rf spectroscopy even at $T_{\text{D}}\,=\,$1\,s.

\section{Ramsey spectroscopy with the UR10 sequence at second-long coherence time\label{sec:LLI section}}
An example of useful implementation of the UR10 sequence is for a test of LLI that is based on a search for a $m_{J}^{2}$-dependent energy shift between the sublevels in the $^{2}F_{7/2}$ manifold that oscillates at frequencies related to Earth's rotation~\cite{Dreissen2022,Shaniv2018}. Such an energy shift is transferred via the UR10 composite pulse Ramsey sequence into modulations in the retrieved population in the $|F,-1/2\rangle$ state after the final rf Ramsey $\pi/2$-pulse. For implementation of the UR10 sequence in the $^{2}F_{7/2}$ manifold, the experimental ``$^{2}F_{7/2}$" sequence, see figure~\ref{fig:levelscheme}\,(c), is used. After the last Ramsey $\pi/2$-pulse and de-excitation on the E3 transition, the retrieved population $P_{r}$ in the $\ket{F,-1/2}$ state is measured via fluorescence detection. If a phase accumulates during the Ramsey dark time, this population will be reduced.

\begin{figure}[ht]
    \centering
    \includegraphics[width=0.6\textwidth]{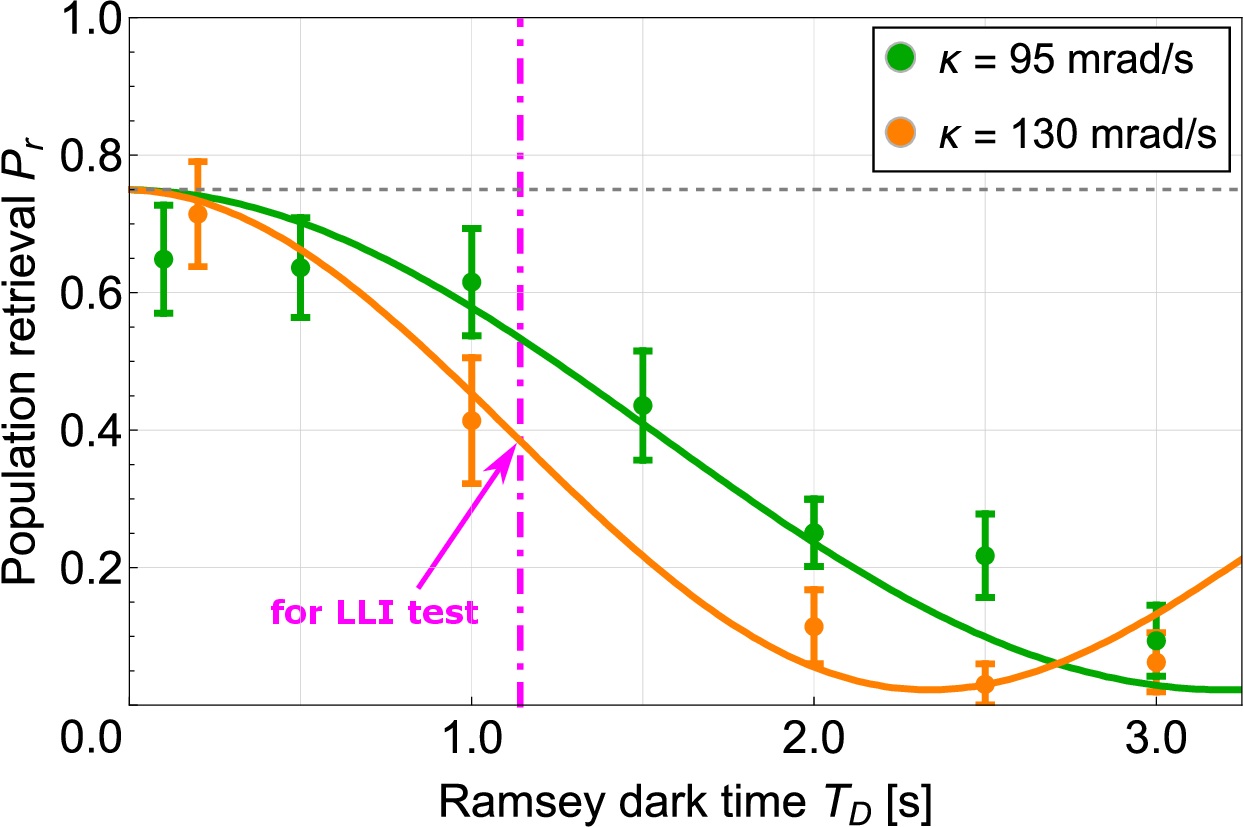}
    \caption{The retrieved population $P_{r}$ as a function of Ramsey dark time $T_{\text{D}}$ at different values of the $\kappa$. The colored points indicate measured data, and the curves are obtained from simulations. The orange and green curves correspond to $\kappa\,=\,130$\,mrad/s and $\kappa\,=\,95$\,mrad/s with an axial secular frequency of $\nu_{\text{ax}}\,=\,237$\,kHz and $\nu_{\text{ax}}\,=\,202$\,kHz, respectively. Here, the quadrupole moment $\Theta$ of the $^{2}F_{7/2}$ state was taken to be $\Theta\,=\,-0.0297(5)\,ea^{2}_{0}$~\cite{Lange2020coherent}. The grey dashed line at 75\,\% represents the maximum contrast of the experiment. The contrast is limited by the excitation and de-excitation probability on the E3 transition of 85\,\%. The Ramsey dark time of $T_{D}\,=\,1.15$\,s, as used is our work in reference~\cite{Dreissen2022}, is indicated with the vertical dash-dotted purple line.}
    \label{fig:Ramseydarktimescan}
\end{figure}

Other than a potential Lorentz violation, the quadratic term also contains the quadrupole shift (see section~\ref{sec:theory}), which is induced by the electric field gradient in the linear rf Paul trap. From both $m_{J}^2$-dependent energy shifts a phase accumulates during the Ramsey dark time $T_{\text{D}}$. However, the accumulated phase caused by a hypothetical violation of the LLI will oscillate specifically at the sidereal day frequency ($\omega_{\oplus}\,=\,2\pi/23.934$\,h) and its harmonics~\cite{Dreissen2022}. To get the highest sensitivity to these types of oscillations, $T_{\text{D}}$ is set such that $|dP_{r}/dT_{\text{D}}|$ is largest. To find this value, $P_{r}$ is measured as a function of $T_{\text{D}}$, as shown in figure~\ref{fig:Ramseydarktimescan}. The different colors correspond to measurements at different values of $\kappa$. The value of $\kappa$ is modified via the axial trap confinement. The data points show the experimental results and the curves are obtained from simulations. The orange and green curves correspond to $\kappa\,=\,130$\,mrad/s and $\kappa\,=\,95$\,mrad/s with an axial secular frequency of $\nu_{\text{ax}}\,=\,237$\,kHz and $\nu_{\text{ax}}\,=\,202$\,kHz, respectively. Both simulated curves agree well with the measured data, for dark times of up to $T_{D}\,=\,2.5$\,s. The observed coherence time is a factor of $10^{4}$ longer compared to a simple Ramsey sequence, i.e., two $\pi/2$-pulses, in which case we measured a coherence time of $t_{\text{coh}}\,\approx\,250$\,µs. The grey dashed line at 75\,\% represents the maximum contrast of the experiment. The contrast is limited by the excitation and de-excitation probability on the E3 transition of 85\,\%. The optimum Ramsey dark time for a test of LLI was determined to be $T_{D}\,=\,1.15$\,s for $\kappa\,=\,130$\,mrad/s, as indicated by the vertical dash-dotted pink line in the figure. At these operating conditions, we have demonstrated an unprecedented sensitivity to a potential LLI violation of $\sigma_{\kappa_{\text{LV}}}\,=\,372(9)/\sqrt{\tau}$\,mrad$\cdot$s$^{-1}$~\cite{Dreissen2022}, enabling 10 times shorter averaging times compared to the previous best test of LLI in Yb$^{+}$~\cite{Sanner2019}. Here, the quadrupole moment $\Theta$ of the $^{2}F_{7/2}$ state was taken to be $\Theta\,=\,-0.0297(5)\,ea^{2}_{0}$~\cite{Lange2020coherent}, where $e$ is the electron charge and $a_{0}$ is the Bohr's radius.

Alternatively, we can also determine the quadrupole shift and extract the quadrupole moment of the $^{2}F_{7/2}$ state. The relation of $\kappa$ and $\nu_{\text{ax}}$ is given as~\cite{Shaniv2016atomic}
\begin{equation}
    \kappa\,=\,-2\pi\cdot\frac{1}{4h}\cdot\frac{J(1+J)-3m_{J}^{2}}{J(2J-1)}\cdot\frac{dE_{z}}{dz} \cdot\Theta\cdot(3\cos^{2}\beta-1)\,\text{,}
    \label{equ:kappa_nu}
\end{equation}
where $h$ is the Planck's constant and $\beta$ is the angle between the trap axis and the quantization magnetic field. In our case, $J\,=\,7/2$, $m_{J}\,=\,1/2$, $\beta\,=\,26.8(4.0)$\,$^{\circ}$ and $dE_{z}/dz$ is the electric field gradient that can be determined from the axial secular frequency according to $dE_{z}/dz\,=\,m_{\text{Yb}}\cdot(2\pi\nu_{\text{ax}})^{2}/e$, as contributions of the field gradient from other sources are much lower. Here, $m_{\text{Yb}}$ is the mass of ytterbium ion. From a fit of the data at a secular frequency of $\nu_{\text{ax}}\,=\,235.72(10)$\,kHz, we obtain a fitted value of $\kappa\,=\,$123(5)\,mrad/s. With these parameters, we can determine the quadrupole moment of the $^{2}F_{7/2}$ state as $\Theta\,=\,-0.0298(38)\,ea^{2}_{0}$. This value agrees with the value deduced from clock measurements at $10^{-18}$ level in reference~\cite{Lange2020coherent} of $\Theta\,=\,-0.0297(5)\,ea^{2}_{0}$. At this level, the uncertainty of $\Theta$ is governed by the uncertainty of the orientation of the magnetic field $\sigma_{\beta}\,=\,4$\,$^{\circ}$. At the current value of $\beta$, $\Theta$ is most sensitive to $\sigma_{\beta}$. By choosing $\beta\,=\,0$, we can sufficiently suppress the effect from $\sigma_{\beta}$ by maximizing $(3\cos^{2}\beta-1)$. At this extreme point ($\beta\,=\,0$), the same uncertainty of $\sigma_{\beta}\,=\,4$\,$^{\circ}$ and $\sigma_{\kappa}\,=\,5$\,mrad/s already leads to an uncertainty of $\Theta$ as $\sigma_{\Theta}\,=\,8\times10^{-4}\,ea^{2}_{0}$.

\section{Scalability of the rf sequence for multi-ion spectroscopy}\label{multi-ion}
For more sensitive Ramsey spectroscopy based on the composite rf pulse sequence with multiple ions, we need to determine the homogeneity of the quantization B-field and the driving rf-field. This is done via measurements of the center frequencies of the $\ket{S,-1/2}\,\rightarrow\,\ket{D,-5/2}$ E2 transition and the rf Rabi frequencies in the spin-$1/2$ $^{2}S_{1/2}$ ground state along an ion Coulomb crystal. From the spread of the center frequencies of the E2 transition and the spread of the rf Rabi frequencies along the ions, we can deduce the B-field gradient and the rf-field gradient, respectively.

\begin{figure}[ht]
    \centering
    \includegraphics[width=0.8\textwidth]{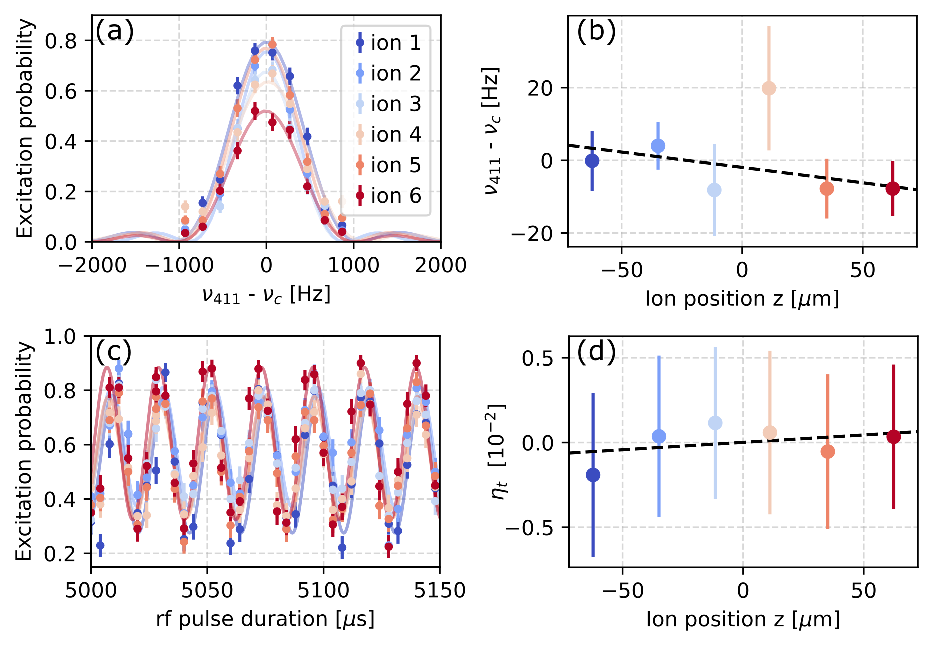}
    \caption{Homogeneity measurement of the quantization magnetic field and the driving rf field for rf spectroscopy. All measurements are done with simultaneous spectroscopy of six ions along a 120\,µm Coulomb crystal. (a) The measured center frequency of the E2 transition. Frequencies are subtracted by mean frequency of the E2 transition $\nu_{c}$. (b) The E2 transition frequency of each ion at different positions of the Coulomb crystal. From a linear fit, the gradient of the center frequency gradient is determined to be $d\nu/dz\,=\,$\,-0.085(61)\,Hz/µm, corresponding to a B-field gradient of $dB/dz\,=\,$\,-3.0\,pT/µm. The maximum observed difference in the center frequency is $\delta\nu\,=\,$10.2\,Hz, corresponding to $\Delta\eta_{\omega}\,=\,0.0004$ for rf spectroscopy in the $^{2}F_{7/2}$ manifold. (c) The measurement of the population transfer of the ions within the $^{2}S_{1/2}$ electronic ground state with respect to the rf pulse duration from 5\,ms to 5.15\,ms. (d) The extracted $\eta_{t}$ and the fitted gradient of $d\eta_{t}/dz\,=\,9(11)\times10^{-6}$\,/µm, corresponding to $\Delta\eta_{t}\,=\,0.001$.}
    \label{fig:homogeneity plots}
\end{figure}

The measurement is done via simultaneous spectroscopy of six ions along a 120\,µm Coulomb crystal. Each ion is represented with a different color. Figure~\ref{fig:homogeneity plots}\,(a) shows the excitation probability of the $\ket{S,-1/2}\,\rightarrow\,\ket{D,-5/2}$ transition. The variation in the observed excitation probability is caused by the finite waist of the 411\,nm laser beam of $w_{411}\,=\,83$\,µm as described in section~\ref{subsec:experimental setup}. Figure~\ref{fig:homogeneity plots}\,(b) shows the differences of the extracted center frequency compared to a common mean frequency of the E2 transition $\nu_{c}$. From a linear fit, the gradient of the center frequency is determined to be $d\nu/dz\,=\,$\,-0.085(61)\,Hz/µm, corresponding to a B-field gradient of $dB/dz\,=\,$\,-3.0\,pT/µm. The maximum observed difference in the center frequency is $\delta\nu\,=\,$10.2\,Hz, which translates to $\Delta\eta_{\omega}\,=\,0.0004$ for rf spectroscopy in the $^{2}F_{7/2}$ manifold. Figure~\ref{fig:homogeneity plots}\,(c) shows the measurement of the population transfer of the ions within the $^{2}S_{1/2}$ electronic ground state with respect to the rf pulse duration. The population is optically pumped to the $\ket{S,-1/2}$ state and state selective fluorescence detection via the $\ket{S,-1/2}\,\rightarrow\,\ket{D,-5/2}$ E2 transition is applied. To ensure that the state detection is not dependent on the probe light intensities seen by each individual ion, we apply a long 15\,ms laser pulse to transfer the population out of the electronic ground state. Due to the lifetime of the state of about 7\,ms~\cite{Taylor1997investigation,Yu2000lifetime,Tan2021precision}, inefficient shelving is observed in the measurement. Figure~\ref{fig:homogeneity plots}\,(d) shows the extracted $\eta_{t}$ for each ion and a fitted gradient of $d\eta_{t}/dz\,=\,9(11)\times10^{-6}$\,/µm. This gives a maximum of $\Delta\eta_{t}\,=\,0.001$ along the crystal.

The extracted values of $\Delta\eta_{\omega,t}$ over a range of 120\,µm are small compared to the allowed range of $|\eta_{\omega}|\,<\,0.021$ and $|\eta_{t}|\,<\,0.004$. Therefore, multi-ion rf Ramsey spectroscopy with on the order of 10 ions in a linear chain simultaneously could be done with the UR10 sequence to improve the sensitivity of the measurement by $\sqrt{10}$ and perform a better test of LLI in the future.

\section{Conclusion}
We investigated rf composite pulse Ramsey spectroscopy in the meta-stable eight-level $^{2}F_{7/2}$ manifold of a trapped $^{172}$Yb$^{+}$ ion and implemented the robust the UR10 sequence for a sensitive test of LLI~\cite{Dreissen2022}. A potential Lorentz violation would manifest itself as a modulation of the energy difference between Zeeman sublevels at a period related to a sidereal day (23.934\,h). Therefore, an accurate test of LLI requires continuous data taking over the course of several weeks, throughout which the experiment needs to operate stably and reliably. We investigated the robustness of the GSE and UR10 sequences against pulse errors, specifically rf detuning and pulse duration errors, that accumulate during the measurement sequence. Due to the slow population transfer via the E3 transition to the $^{2}F_{7/2}$ manifold, we first characterize the robustness of the rf sequences in the spin-$1/2$ $^{2}S_{1/2}$ ground state of $^{172}$Yb$^+$. 

The stability diagram of two specific sequences, the GSE and the UR10 sequence, were simulated in a spin-$1/2$ system as a function of pulse errors. The final state was compared to the desired state to determine the fidelity ($\mathcal{F}$) of the sequences. We quantitatively investigate the robustness of a sequence to pulse errors by comparing them to our experimental uncertainties of $\delta\eta_{\omega}\,=\,\pm0.021$ and $\delta\eta_{t}\,=\,\pm0.004$, where $\eta_{\omega}$ and $\eta_{t}$ are a measure for the detuning and the pulse duration error, respectively. We simulated the sequences at a Ramsey dark time of $T_{\text{D}}\,=\,$10\,ms (in total 50 rephasing pulses) and find good agreement with experimental results. Under these conditions, we found that the GSE sequence performs just within the limit of our required stability, while the UR10 sequence has 38 and 13 times higher tolerance to pulse duration errors and detuning errors, respectively. The simulations were extended to an eight-level quantum system and the robustness of the rf sequences was investigated in the $^{2}F_{7/2}$ manifold. The stability diagrams show that the UR10 sequence only introduces an error of $\epsilon_{\text{UR10}}\,=\,3\times10^{-10}$ even at $T_{\text{D}}\,=\ $1\,s.

The UR10 sequence was experimentally implemented in the $^{2}F_{7/2}$ manifold and a coherent signal of up to $T_{\text{D}}\,=\ $2.5\,s was observed. For a test of LLI, optimal measurement conditions were found at a Ramsey dark time of $T_{\text{D}}\,=\ $1.15\,s and the highest sensitivity to the validity of LLI in the electron-photon sector to date was reached~\cite{Dreissen2022}. Owing to the robustness of the rf sequence to pulse errors, a further extension of the method to larger ion Coulomb crystals is possible for more accurate tests of LLI in the future. Initial measurements show that the gradient of the quantization magnetic field and rf field meet the requirements to extend the UR10 sequence to a linear Coulomb crystal of 10 ions extended over 120\,µm without a significant loss of fidelity from ion to ion. We also showed that this method can be used to accurately measure quadrupole moments of meta-stable states, such as that of the $^{2}F_{7/2}$ state. With our measurement, we determined the quadrupole moment of the $^{2}F_{7/2}$ state to be $\Theta\,=\,-0.0298(38)\,ea^{2}_{0}$. With a rotated axis of the magnetic field ($\beta\,=\,0$), an uncertainty of $\sigma_{\Theta}\,=\,8\times10^{-4}\,ea^{2}_{0}$ can be achieved with this method, which is competitive to optical clock measurements at the $10^{-18}$ level~\cite{Lange2020coherent}

\section*{Acknowledgments}
We would like to thank Melina Filzinger, Nils Huntemann and Ekkehard Peik for helpful discussions.

This project has been funded by the Deutsche Forschungsgemeinschaft (DFG, German Research Foundation) under Germany’s Excellence Strategy – EXC-2123 QuantumFrontiers –390837967 (RU B06) and through Grant No. CRC 1227 (DQ-mat, project B03). This work has been supported by the Max-Planck-RIKEN-PTB-Center for Time, Constants and Fundamental Symmetries. L.S.D. acknowledges support from the Alexander von Humboldt foundation.

\section*{References}
\bibliographystyle{iopart-num.bst}
\bibliography{references.bib}

\clearpage
%\appendix

%\section{rf coil configuration\label{ap:rf_coil}}
%The coil for rf spectroscopy consists of an inner coil of 19 windings to couple the rf power in and an outer coil of 27 windings, which delivers the rf power to the ion. The wire used for both coils has a diameter of 1\,mm and the outer coil has a diameter of 4.5\,cm resulting in a self-capacitance of 77.7\,pF. A tunable capacitor of 1-2\,pF, an inductor of $L\,=\,2.66\times10^{-5}$\,H and a resistor of $R\,=\,100\,\Omega$ are connected to the outer coil to tune the coil's resonance. The Q factor of the rf coil is 18.6 and the ring down time is 7\,µs. 

\end{document}